\documentclass[aps,showpacs,epsf,twocolumn,prb]{revtex4}
\usepackage{colordvi}
\usepackage{amssymb}
\usepackage{amsmath}
\usepackage{epsf}
\usepackage{mathrsfs}
\usepackage[dvips]{graphics}
\usepackage[dvips,usenames]{color}
\begin{document}
%\draft
\def \beq{\begin{equation}}
\def \eeq{\end{equation}}
\def \bea{\begin{eqnarray}}
\def \eea{\end{eqnarray}}
\def \bem{\begin{displaymath}}
\def \eem{\end{displaymath}}
\def \P{\Psi}
\def \Pd{|\Psi(\boldsymbol{r})|}
\def \Pds{|\Psi^{\ast}(\boldsymbol{r})|}
\def \Po{\overline{\Psi}}
\def \bs{\boldsymbol}
\def \bl{\bar{\boldsymbol{l}}}
%\begin{document}
\title{Numerical study of one-dimensional and interacting Bose-Einstein condensates in a random potential}
\author{Eric Akkermans$^{1}$, Sankalpa Ghosh$^{1,3,4}$ and Ziad Musslimani$^{2}$}
\affiliation{$^{1}$Department of Physics, Technion Israel Institute of
Technology,
32000 Haifa, Israel\\
$^{2}$Department of Mathematics, Florida State University, Tallahassee, FL 32306-451 \\
$^3$Physics Department, Okayama University, Okayama 700-8530, Japan\\
$^4$Physics Department, Indian Institute of Technology, Delhi, New Delhi
110016, India}
\date{\today}

\begin{abstract}
We present a detailed numerical study of the effect of a disordered  
potential on a confined one-dimensional Bose-Einstein
condensate, in the framework of a mean-field description. For repulsive interactions, we consider the Thomas-Fermi and Gaussian limits and for attractive interactions the behavior of soliton solutions. We find that the disorder average spatial extension  of the stationary density profile  
decreases with an increasing strength of the disordered potential both 
for repulsive and attractive interactions among bosons.  In the Thomas 
Fermi limit, the suppression of transport is accompanied by a strong 
localization of the bosons around the state $k=0$ in momentum space. The time dependent  density profiles differ considerably in the cases we have considered. For attractive Bose-Einstein condensates, a bright
soliton exists with an overall unchanged shape, but a disorder dependent 
width. For weak disorder, the soliton moves on and for a stronger 
disorder, it bounces back and forth between high potential barriers.
\end{abstract}
\date{\today}

\maketitle
\section{Introduction}

The spatial behavior of a wave submitted to a strong enough random potential remains one of the major and still unsolved issues in physics. It is an ubiquitous problem that shows up in almost all fields ranging from astrophysics to atomic physics. The interference induced spatial localization of a wave due to random multiple scattering  has been predicted and named after Anderson \cite{Ander58}. The Anderson localization problem despite its relatively easy formulation has  not yet been solved analytically and still rises a lot of interest. Strong Anderson localization of waves has been observed in various systems of low spatial dimensionality where the effect of disorder is expected to be the strongest \cite{parodi,kergomard,amaynard}. Above two dimensions, a phase transition is expected to take place between a delocalized phase that corresponds to spatially extended solutions of the wave equation and a localized phase that corresponds to spatially localized solutions. The description of  this transition is mainly based on an elegant scaling formulation proposed by Anderson and coworkers \cite{Gang479}. Due to its indisputable importance, the localization of light is a hotly debated but still unsolved 
problem \cite{genack,lag,maret}. The weak localization regime, a precursor of Anderson localization for weak disorder, has been studied in detail both theoretically and experimentally for a large variety of waves and types of disorder \cite{review1,review2,am,josaam}.  

In contrast, relatively little attention has been paid to the extension of Anderson localization to a non-linear medium. Though analytical\cite{DS86,DR87,KIV90} as well as numerical work
have been done to address this issue, no clear-cut answers have
been obtained to ascertain how localization is affected by the
presence of a non-linear term in a Schr\"odinger type wave
equation. This is the purpose of this paper to address this issue in the context of the behavior of a one-dimensional Bose-Einstein condensate
(BEC) in the presence of a disordered optical potential, since it has raised recently a great deal of interest \cite{disorder1,disorder1a,disorder2,disorder3,Dam03,sch05,
SKSZL04,Wang04,GC04,PVC05,NBNP05,fallani,modugno}. Transport of a magnetically trapped BEC above a corrugated microship has been theoretically studied recently \cite{paul}. The possibility of tuning
random on-site interaction has also been considered
\cite{GWSS05}. Using Feshbach resonances, it is possible to
switch off the interaction among bosons which will then be allowed
to propagate through a set of static impurities created by other
species of atom. This may lead to an experimental realization of
the Anderson localization transition. The corresponding theoretical model  has 
been proposed and analyzed \cite{GC04, PVC05} for one-dimensional systems, {\it i.e.} in the absence of transition. The
other issue is to understand the interplay of interaction induced
non-linearity and disorder on the Bose-Einstein condensate. One-dimensional systems are especially interesting since the effect of disorder is the strongest and such systems are experimentally realizable. Experiments in this direction have been
performed recently \cite{disorder1,disorder1a,disorder3} which show a
suppression of the expansion of the BEC cloud once it is released from the trap.

In this paper we present a numerical study of 
the effect of a disordered potential on one-dimensional condensates with either attractive or repulsive interaction in the framework of the mean-field approximation and compare between these two cases.  Studies of the propagation of a quasi one-dimensional 
BEC in a disordered potential have been carried out mostly in the repulsive Thomas-Fermi
limit \cite{disorder1, disorder1a, disorder3, sch05, modugno}.  We also consider this limit and we find numerical 
evidence 
that the suppression of the BEC expansion after the release from the 
trap, is due to localization in momentum space around the 
state 
$k=0$, with becomes stronger for an increasing strength of disorder. This suggests that the momentum 
spectroscopy of disordered quasi one-dimensional  BEC may give important information about 
its transport properties. In addition, we consider the Gaussian limit of a strong confinement and the bright soliton solution for an attractive interaction. We also propose a model for the disorder where both the strength and the harmonic content can be independently varied. 
\medskip

The comparison between the different cases is
motivated by the fact that depending on the strength and sign of
the effective interaction among bosons in an effectively one-dimensional
BEC, various types of scenarios may be realized. The interplay
between these different types of interaction and disorder should
lead to different types of stationary as well as time-dependent
behavior of the density profile. We consider three such regimes that cover both the
repulsive and attractive interaction and where the system can indeed be
well described within the mean-field approximation. The corresponding mean-field
is given by the Gross-Pitaevskii equation with modified
coupling constant \cite{Olshanii98} (in comparison to the three dimensional case) and it takes 
the form of a non-linear Schr\"odinger equation.
Its solutions in the absence of disorder have been thoroughly 
studied \cite{PSW00,soliton02,CC02,book1}. We employ a numerical scheme
\cite{P76, AM03, AM04, AM05} which has been recently developed to study stationary solutions of
this non-linear Schr\"odinger equation in the presence of a disordered potential. The scheme is based
on a rapidly converging spectral method. Then we look at the time evolution
of the stationary  profile after switching off the trap potential.
Subsequently, we analyze our
solutions and compare them to those obtained in the
absence of disorder. Our study unveils an interesting picture for 
the interplay between the nature and strength of interaction and
a random potential.

The organization of the paper is as follows. In
section II we briefly review the stationary density profiles of
an effectively one-dimensional BEC in the absence of disorder and  in the mean field regime.
Then, in section III, we introduce our numerical scheme and we define our model of disorder 
on such one-dimensional condensates. In section IV, we present our numerical results for the Thomas-Fermi limit. In section IV-A we compare our results with recent works
on this subject \cite{disorder1, disorder1a, disorder3, sch05, modugno,paul}. In section V, the effect of disorder in the confinement dominated Gaussian regime is discussed.
Both sections pertain to the situation of repulsive interaction among bosons. In section VI, we
discuss the effect of disorder on a bright solitonic condensate which corresponds to an attractive  effective
interaction. In the last section we summarize and present  the general conclusions derived from
our results.

\section{Stationary solutions in the absence of disorder}

\subsection{One-dimensional repulsive Bose-Einstein condensate in a trap}

We review briefly the mean field description  of a quasi one-dimensional Bose gas  with short range repulsive interaction, in a
cylindrical harmonic trap along the $z$-axis, and in the absence of disorder.
Details are given in references \cite{pita, PSW00}. The
Gross-Pitaevskii equation provides a mean field description of
the three dimensional interacting gas and it is given by \bea i\hbar \frac{\partial
\Psi}{\partial t}&=& -\frac{\hbar^2}{2m} \nabla^2 \Psi+
\frac{1}{2}(m\omega_z^2 z^2+ m\omega_{\perp}^2
(x^2+y^2))\Psi \nonumber \\
& & \mbox{}+ \frac{4 \pi a \hbar^2}{m}|\Psi|^2 \Psi
\label{nlse1}\eea where $\omega_{z}$ and $\omega_{\perp}$ are
respectively the harmonic trap frequencies along the $z$-axis and along
the radial direction; $a_{z}=\sqrt{\frac{\hbar}{m\omega_{z}}}$ and
$a_{\perp}=\sqrt{\frac{\hbar}{m\omega_{\perp}}}$ are the
corresponding harmonic oscillator length scales. The interaction
is characterized by the $s$-wave scattering length $a$,
that is positive for a repulsive interaction. For tight trapping
conditions ($\omega_z \ll \omega_{\perp}$), all atoms are in the
ground state of the harmonic trap in the radial direction and the
condensate is effectively one-dimensional. Nevertheless, for
$a_{\perp}
> a$, the effective coupling constant along the $z$-direction is still characterized
by $a$ and it is given by $g_{1d}=2a\hbar \omega_{\perp}$
\cite{Olshanii98}. The corresponding mean-field behavior is governed 
by the Gross-Pitaevskii equation,
 \beq i\hbar \frac{\partial \Psi}{\partial t}=
-\frac{\hbar^2}{2m}\frac{\partial^2 \Psi}{\partial z^2}+
\frac{1}{2}m\omega_z^2 z^2\Psi + g_{1d} |\Psi|^2
\Psi \label{nlse1}\eeq where $\Psi$ is the condensate
wavefunction along the $z$-axis. We look for stationary solutions of the form $\Psi(z,t)
=\phi(z)\exp(-i {\tilde \mu} t)$ where $\tilde \mu$ is the chemical potential. The
corresponding one-dimensional density is  $\rho_{1d}= |\phi (z)|^2$. The interaction strength may be expressed  in terms of the dimensionless coupling constant $\gamma$,
 \beq \gamma=\frac{mg_{1d}}{\hbar^2 \rho_{1d}}, \eeq
which is the ratio of the  mean-field interaction energy density
to the kinetic energy density. For $ \gamma\ll 1$, the
gas is weakly interacting and, in contrast to higher space
dimensionalities, in one-dimension the gas can be made strongly
interacting by lowering its density. For larger values of the interaction strength
$\gamma$, the Gross-Pitaevskii equation
(\ref{nlse1}) does not provide anymore a correct description, the gas enters into the Tonks-Girardeau regime
\cite{TG36} and behaves like free fermions.

Starting from (\ref{nlse1}), a dimensionless form can be achieved that is given by
\beq i
\partial_t \Psi + \partial_{z} ^2 \Psi -z^2 \Psi -2\alpha_{1d}|\Psi|^2
\Psi=0 \, \ , \label{tdimnlse}\eeq where use has been made of rescaled
length and time, $z \rightarrow \frac{z}{a_{z}}$, $ t  \rightarrow
\frac{\omega_{z}}{2}t$ and $\Psi \rightarrow \sqrt{a_z} \Psi$. The
dimensionless parameter $\alpha_{1d}$, or equivalently the
coherence length $\xi$,  is defined by  \beq \alpha_{1d}=
\frac{2aa_z}{a_{\perp}^2} = {1 \over 2 \xi^2} \, ,
\label{alpha1d} \eeq and it accounts for both interaction and confinement. By rescaling the chemical potential, $\mu
\rightarrow \frac{\tilde \mu}{\hbar \omega_z}$, we obtain for the
time-independent Gross-Pitaevskii equation the expression 
 \beq \mu \phi + \partial_{z} ^2 \phi -z^2 \phi -2\alpha_{1d}|\phi|^2
\phi=0 \, \ . \label{dimnlse}\eeq Henceforth we shall express our results in
terms of these dimensionless quantities unless otherwise specified.
We mention now two limiting regimes of interest that can be described by Eq.(\ref{dimnlse}).

\subsubsection{Thomas-Fermi  limit}

 For a chemical potential $\tilde \mu$ larger than the level spacing, namely for ${\tilde \mu} \gg \hbar \omega_z$ ({\it i.e.} in dimensionless units $\mu \gg 1$),  the gas is in the Thomas-Fermi  regime. Thus the kinetic energy term becomes
negligible.  We denote by $\rho_{TF}$ and $\mu_{TF}$ the corresponding condensate density and chemical potential. We have 
\beq \rho_{TF} =\frac{\mu_{TF} -
z^2}{2\alpha_{1d}} \,  \Theta(\mu_{TF}-z^2) \label{TFdensity} \, \ . \eeq
The number of bosons is given by $N=\int_0^{L_{TF}} dz \rho_{TF}$, where $L_{TF}=\sqrt{\mu_{TF}}$ is the Thomas-Fermi  length. Eliminating $L_{TF}$, we obtain,
 \beq
\mu_{TF}= \left(
\frac{3N\alpha_{1d}}{4 \sqrt{2}} \right)^{2/3}. \eeq

\subsubsection{Gaussian limit}

The other limit ${\tilde \mu} \ll \hbar \omega_{z}$, corresponds to a regime where the
single particle energy spacing is larger than the interaction
energy so that the gas behaves like $N$ bosons in a harmonic trap potential. Thus, we have an ideal gas condensate with a Gaussian density profile.

As we shall see later, the effect of a disordered potential on the 
condensate dynamics for both limiting cases is significantly different. 

\subsection{ One-dimensional attractive Bose-Einstein condensate}

We also consider the case where the $s$-wave scattering length $a$ is negative.  The effective
interaction among bosons is thus attractive. This situation can also be described by means of Eqs.(\ref{nlse1}-\ref{tdimnlse}). In the absence of confinement and when $\alpha_{1d} = -1$\cite{book1}, Eq.(\ref{tdimnlse}) admits a  moving bright soliton solution of the form
\beq
\Psi(z,t)=\sqrt{\mu} \, \frac{\exp \left( i(\frac{V_s}{2}z+(\mu-{V_s^2 \over 4})t+\phi_0) \right)}
{\cosh \left( \sqrt{\mu} (z -V_s t -z_0) \right)}
\label{boost}
\eeq
 where $V_s$ and $\mu > 0$ are respectively the velocity and the chemical potential of the soliton solution and
($z_0$, $\phi_0$) refer to the translational and global phase
invariance of Eq.(\ref{tdimnlse}). In particular, if $V_s = 0$ and choosing for simplicity the gauge $z_0 = \phi_0 =0$, then $\Psi (z,t) = \phi (z) \exp (i \mu t)$ with 
\beq \phi(z)=\sqrt{\mu} \, \
\mbox{sech}(\sqrt{\mu} \, \ z) \, \ , \label{ANLSE} \eeq
which satisfies the time-independent nonlinear Schr\"odinger equation
\beq -\mu \phi(z)+\partial_z^2
\phi(z)+2|\phi|^2\phi=0 \,  . \label{ANSEE1} \eeq
The chemical
potential $\mu$  is proportional to the square of the inverse
width of the soliton. Such a soliton has been experimentally
observed \cite{soliton02} and theoretically studied \cite{CC02}
for cold atomic gases. 

\section{Numerical method for disorder and non-linearity}

\subsection{Spectral  method}

We start by considering the dimensionless time-independent Gross-Pitaevskii
equation
 \beq \mu \phi  +\partial_z^2 \phi -z^2\phi(z)-V_d (z)\phi - 2\alpha_{1d}|\phi|^2\phi=0\;,
\label{gnlse}
\eeq
in the presence of a disorder potential $V_d(z)$. Upon discretization, this potential is defined at each site of a lattice and it is given by the product of a constant strength $V_{m}$ times a random number $\omega$ which is uniformly distributed between $0$ and $1$. Using a Gaussian approximation with mean $\sigma$ (the lattice spacing), the disorder potential can be written as a continuous function
 \beq
V_d(z, \omega)=\omega V(z)\;,\eeq with \beq V(z-z')= \lim_{\sigma
\rightarrow 0} V_{m}
\exp\left(-\frac{(z'-z)^2}{\sigma^2}\right) \, \ . \label{disorder}
\eeq
A disorder potential generated in this way
varies rapidly over a length scale of the order of a lattice
spacing.  We wish however to use a smoother potential more appropriate for the description of typical disorders generated in experiments
\cite{disorder1, disorder3}. To that purpose, we consider the discrete random variable $\omega$ defined at each lattice site and we discard from its Fourier spectrum all wavenumbers that are above a given cutoff $k_c = 2 \pi / \lambda_c$.  The inverse Fourier transform $\omega (\lambda_c) = \omega^c$ provides a random potential that varies on length scales larger or equal to $\lambda_c$ and which can be formally written as
%-----------------------------------------------
\beq V^c (z) =  V_m \int dk e^{ikz}\left[
e^{-\left({\frac{k}{k_c}}\right)^M} \int d\zeta \, \ \omega (\zeta) e^{-ik\zeta}\right]\; ,
\eeq
%-----------------------------------------------
where $M$ is a large enough number. The new random variable $\omega^c (\lambda_c, z)$ thus generated is different from $\omega$. Whereas the average value of $\omega$ is, by definition, equals to $1/2$, we obtain, for example,  that for $k_c = 6$, the average value of $\omega^c$ is about $2 \times  10^{-2}$. Typical examples of such slowly varying potentials obtained by changing $\lambda_c$ are given in section V in Fig.\ref{fig:Gauss-den-time2}({\it a,c,e}). The disorder potential $V^c  = V_m \omega^c (\lambda_c,z)$ that we consider is thus characterized by two quantities: its strength $V_m$ and the scale $\lambda_c$ of its spatial variations.  Eq.(\ref{gnlse}) rewrites
%-----------------------------------------------
\beq \mu
\phi_{\omega}  +\partial_z^2 \phi_{\omega} -z^2\phi_{\omega}-
V_m \omega^c (\lambda_c,z)\phi_{\omega} -
2\alpha_{1d}|\phi_{\omega}|^2\phi_{\omega}=0\; . \label{gnlse2}
\eeq
%-----------------------------------------------
The local density  for a given realization of disorder is $\rho_\omega (z) = |\phi_\omega (z)|^2$ and the number $N$ of bosons is determined by the condition $N = \int dz \, \ \rho_\omega (z)$. By direct inspection of the different terms that show up in Eq.(\ref{gnlse2}), we see that  disorder effects are obtained either by comparing them to interactions, {\it i.e.}, by comparing the disorder length scale $\lambda_c$ to the coherence length $\xi$ defined in (\ref{alpha1d}). If the  ratio $\lambda_c / \xi$ is small, disorder is strongly varying spatially and its effect overcomes that of interactions. We also compare the effective disorder strength $V_m  \omega^c$ to the chemical potential $\mu$. This can be achieved by defining the local dimensionless random variable
\beq
s= {V_m \over \mu}  \omega^c \, \ .
\label{parameters}
\eeq
We will consider its average over  configurations denoted by $\langle s \rangle$.  The parameter $s$ allows to compare between the chemical potential and heights of  barriers of the disorder potential. This parameter, as we shall see, plays also an important role in the study of the time evolution of the density once the trapping potential is released. It controls the spatial extension of the cloud as a function of time.
Finally, we consider boundary conditions for Eq.(\ref{gnlse2}) obtained by demanding that  for a given realization of disorder,
$\phi_{\omega}(z)$ vanishes for $|z|\longrightarrow
+\infty$.

We now turn to the description of
the numerical method used to solve Eq.(\ref{gnlse2}). The fact that  it is
random, makes it very challenging for conventional
numerical schemes to be implemented. The numerical scheme we use here is based on the spectral renormalization method
that has been recently suggested by Ablowitz and Musslimani
\cite{AM05} (see also \cite{AM03,AM04}) as a generalization of the
so-called Petviashvili method \cite{P76}.  Spectral
renormalization is particularly suitable for this type of
problems for its ease to handle randomness. To this end, for a
fixed realization, we define the Fourier transform
%-------------------------------------------------------------------------------------------------
\beq
\hat{\phi}_{\omega}(k)=\mathcal{F}[\phi_{\omega}(z)]=\int dz
\phi_{\omega}(z) e^{-ikz}\;. \eeq
By Fourier
transforming  Eq.(\ref{gnlse2}) we obtain
 \beq
\hat{\phi}_{\omega}(k)=\frac{2\alpha_{1d}\mathcal{F}[|\phi_{\omega}|^2\phi_{\omega}]
+\mathcal{F}[z^2\phi_{\omega}(z)]+\mathcal{F}[V^c(z,\omega
)\phi_{\omega}]} {\mu-k^2}. \label{it1} \eeq
%-------------------------------------------------------------------------------------------------
In general, the
solution of this equation is obtained by a relaxation method or
successive approximation technique where one gives an initial
guess and iterates until convergence is achieved. However, this
relaxation process is unlikely to converge. To prevent this problem, we
introduce a new field variable $\psi_{\omega}(z)$ using a scaling parameter $p_\omega$,
%-------------------------------------------------------------------------------------------------
\beq
\phi_{\omega}(z)= p_\omega \psi_{\omega}(z)\;,\;\;\;\;\;\;\;\;\;
\hat{\phi}_{\omega}(k)=p_\omega \hat{\psi}_{\omega}(k)\;.
\eeq
%-------------------------------------------------------------------------------------------------
Substituting into Eq.(\ref{it1}) and by adding
and subtracting the term $r \hat{\phi}_{\omega}(k)$ (with $r>0$) to avoid
division by zero, we obtain the following scheme

%-------------------------------------------------------------------------------------------------
\begin{eqnarray}
\label{it2}
& & {\hat{\psi}_{\omega}}^{(m+1)}(k) = \left(\frac{r+\mu}{r+
k^2}\right)\hat{\psi}_{\omega}^{(m)} -\frac{{\cal
F}[z^2\psi_{\omega}^{(m)}]}{r+ k^2} - \nonumber \\
&-& \frac{{\cal F}[V^c (z,\omega)\psi_{\omega}^{(m)}]}{r+
k^2} - 2\alpha_{1d}|p_\omega^{(m)}|^2\frac{{\cal
F}[|\psi_{\omega}^{(m)}|^2\psi_{\omega}^{(m)}]}{r+ k^2}\;,
\label{scheme}
\end{eqnarray}
%-------------------------------------------------------------------------------------------------
where $p_\omega^{(m)}$ are given by the following consistency
condition
 \beq
|p_\omega^{(m)}|^2=\frac{\langle\hat{\psi}_{\omega}^{(m)}, (\mu
-k^2)\hat{\psi}_{\omega}^{(m)}-\mathcal{F}[z^2\psi_{\omega}^{(m)}]-\mathcal{F}[V^c \psi_{\omega}^{(m)}]\rangle}
{\langle
\hat{\psi}_{\omega}^{(m)},\mathcal{F}[|\psi_{\omega}^{(m)}|^2\psi_{\omega}^{(m)}]\rangle}
\label{scheme22} \eeq where the inner product in Fourier space is defined by 
$$\langle \hat{f},\hat{g}\rangle =\int \hat{f}\hat{g}dk \, \ .$$

\subsection{Time dependent evolution}
To describe the time evolution of the
stationary solutions, we use a time splitting Fourier spectral method
that has been described in detail \cite{BJM03}. We describe it briefly with a comment on its limitation.

After switching off the trap, the time evolution is governed by the equation \bea i \partial_t \Psi_{\omega}(z,t) &=&
-\partial_z^2 \Psi_{\omega}(z,t)+ V_m \omega^c (\lambda_c,z)\Psi_{\omega}(z,t) \nonumber \\
& &\mbox{+} 2\alpha_{1d}|\Psi_{\omega}(z,t)|^2\Psi_{\omega}(z,t)
\label{gtnlse2} \, \ , \eea
with $\Psi_{\omega}(z,0)=\phi_{\omega}(z)$.
Equation (\ref{gtnlse2}) is solved
in two distinct steps. We solve first
 \beq i \partial_{t}
\Psi_{\omega}(z,t)=-\partial_z^2 \Psi_{\omega}(z,t), \label{time1}
\eeq for a time step of length $\Delta t$ and then,
 \beq i \partial_t  \Psi_{\omega}(z,t)= V_m \omega^c
(\lambda_c,z)\Psi_{\omega}(z,t) +
2\alpha_{1d}|\Psi_{\omega}(z,t)|^2\Psi_{\omega}(z,t)
,\label{time2} \eeq for the same time step. The first of these two
equations, (\ref{time1}), is discretized in space by the Fourier
spectral method and time integrated. The solution is then used
as the initial condition for the second equation (\ref{time2}).
The commutator between the two parts of the Hamiltonian that
appears in the right hand side of (\ref{time1}) and
(\ref{time2}) is disregarded in this process. The resulting  
error is significant if this commutator is large compared to other terms in
the equation. This is the case if the disordered potential strongly fluctuates (which is not considered in the present numerical
work). Notice that, by definition, this method ensures the conservation of the total number of particles.

\section{Thomas-Fermi limit}

\subsection{Stationary solutions}

\begin{figure}[ht]
\centerline{ \epsfxsize 10cm \epsfysize 8cm \epsffile{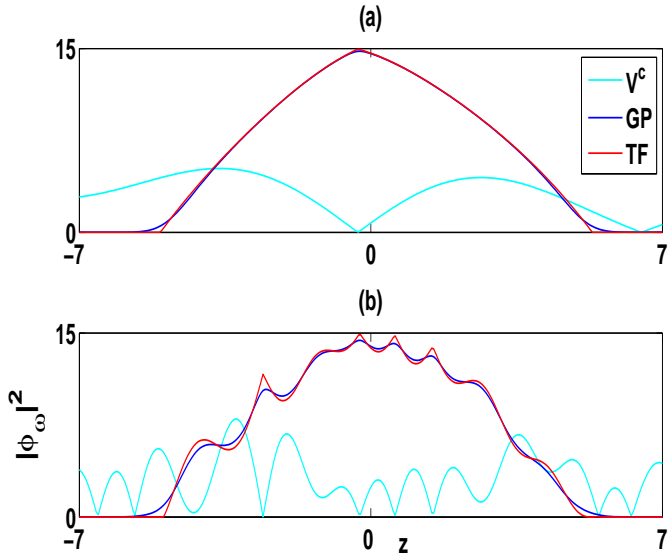} }\caption{\em  (a) Behavior of the condensate density $|\phi_\omega|^2$
obtained from the Thomas-Fermi approximation and from the Gross-Pitaevskii equation for a spatially slowly
varying disordered potential characterized by $\lambda_c \approx 12 \xi$. This corresponds to a weak disorder as compared to interactions. (b) Same plot but for a spatially rapidly
varying disorder such that $\lambda_c \approx 2 \xi$.
The average disorder is kept much below the chemical
potential ($\mu=30$) in both cases, so that $\langle s \rangle \ll 1$. We have taken $\alpha_{1d}=1$
and a number $N$ of bosons equal to $80$. }
\label{fig:TFGPdiff}
\end{figure}

Stationary  solutions to the Gross-Pitaevskii equation (\ref{gnlse}) 
in the presence of a disordered potential and in the Thomas-Fermi limit can be obtained by iterating Eqs. (\ref{scheme}) and (\ref{scheme22}). Then, we compare these solutions with those obtained by directly
considering the Thomas-Fermi approximation in the presence of disorder.
This comparison is displayed on Fig.\ref{fig:TFGPdiff}.

%------------------------------
%------------------------------

Generalizing the Thomas-Fermi approximation (\ref{TFdensity}) so as to include the disorder $V^c$, we obtain for  the corresponding  density the expression,
%-----------------------------------------------
\bea \rho_{TF} (z) &=& \frac{\mu - z^2 - V^c}{2\alpha_{1d}} ,~~ \mu \ge z^2 + V^c \nonumber \\
               &=& 0, ~~\mu <  z^2 + V^c \, . 
\eea
%-----------------------------------------------
It is thus expected that this density presents local maxima and
minima that follow the corresponding ones of the disordered
potential. This trend is indeed very apparent in
Fig.\ref{fig:TFGPdiff}. In  Fig.1.{\it (a)}, we consider a very smooth
disorder such that $\lambda_c$ is much larger than the
coherence length $\xi$ and we observe that apart from  little
deviations, the densities obtained from
the Gross-Pitaevskii equation and from the Thomas-Fermi approximation
match almost exactly as expected. Such a smooth disorder corresponds
to the typical situations encountered in the experiments performed at Orsay\cite{disorder3},  
at LENS\cite{disorder1a} and Hannover \cite{sch05}.
In Fig.1.{\it (b)},
the disorder is stronger {\it i.e.} that it  fluctuates 
on a smaller length scale $\lambda_c$ comparable to $\xi$, thus leading to more local minima and maxima of the disordered potential
within the size of the cloud. The Gross-Pitaevskii density, obtained by solving (\ref{gnlse2}), deviates from the Thomas-Fermi density at these extremal points. Moreover, we observe that the magnitude of those
deviations gets larger when $\lambda_c$ gets smaller, {\it i.e.}, 
for larger spatial variations of disorder. This behavior can be
understood by considering the following expression for the density
$\rho_\omega$
%-----------------------------------------------
\bea \rho_\omega &=& \frac{(\mu -z^2
-V^c (z))}{2\alpha_{1d}}+
\xi^2 \left( \frac{\partial_z ^2 \phi}{\phi} \right) \nonumber \\
&=& \rho_{TF}+ \xi^2 \left( \frac{\partial_z ^2 \phi}{\phi}
\right) \eea
%-----------------------------------------------
which follows straightforwardly from Eq.(\ref{gnlse2}).  In this
expression, the second term in the {\it r.h.s.}, also known as quantum
pressure term, is a correction to the Thomas-Fermi density whose
origin is the zero point motion of the bosons in the condensate.
This correction is proportional to the ratio $(\xi /
\lambda_c)^2$. It becomes larger for a decreasing $\lambda_c$,
namely for a relatively larger effect of interactions driven by
$\xi$. Thus a stronger disorder introduces more appreciable
zero point motion of the bosons so as to reduce the interaction energy
cost. In other words, the behavior of the static
Thomas-Fermi condensate in a random potential is such that the disorder 
potential gets screened by the repulsive interaction \cite{sch05, SR94}.
\medskip

\begin{figure}[ht]
\centerline{ \epsfxsize 7cm \epsfysize 6cm \epsffile{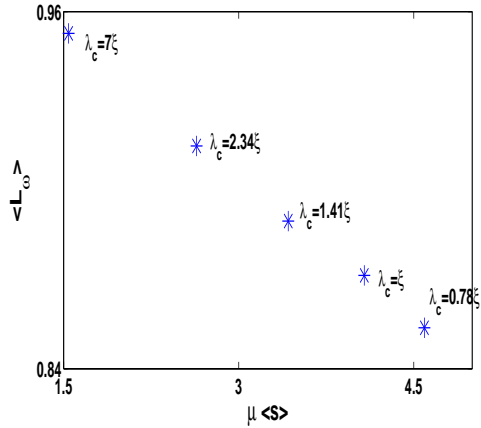} }
%\epsfbox{TFfig_1a.eps}
%\vskip1.0pc
\caption{\em Plot of the width $\langle L_{\omega} \rangle$ averaged over disorder, as a function of
the average strength $\mu \langle s \rangle$. $\langle L_{\omega} \rangle$ is expressed in units of the corresponding extension in the absence of disorder. For a
given value of $\lambda_c$, we average over $200$
realizations of the disordered potential. Such calculations have been done for five different values of $\lambda_c$, the last one being less than the coherence length $\xi$ for which the validity of the
mean-field theory is questionable. We have taken $\mu=30$ and $\alpha_{1d}=1$}
\label{fig:TFwidth}
%%\end{center}
\end{figure}

Another feature of disorder is the spatial extension of the cloud defined, for a given disorder configuration, by
%-----------------------------------------------
\beq L_{\omega} = \sqrt{\overline{z^2}- \overline{z}^2} \label{spatialext}
\eeq
where we have characterized the spatial distribution of the cloud by its moments,
\beq  \overline{z^n} = \frac{\int dz z^n \rho_\omega (z)}{\int dz \rho_\omega (z)} \, \ . \label{rmswidth}
\eeq
%-----------------------------------------------
In Fig.\ref{fig:TFwidth}, we have plotted the configuration average $\langle L_{\omega} \rangle$ of the spatial extension as a function of the average strength $\mu \langle s \rangle$ (see Eq.\ref{parameters}). The average spatial extension of the cloud, in the Thomas-Fermi limit is a decreasing function of the ratio $\lambda_c / \xi$, {\it i.e.}, it decreases when interactions are getting larger than the spatial variation of disorder. We shall see  that this behavior holds true even beyond the Thomas-Fermi approximation.
%--------------------------------------------------------------------------------
%--------------------------------------------------------------------------------

\subsection{Time evolution}
We study now the time evolution of the previous stationary solutions while 
switching off the trapping potential, but keeping the disordered
potential. 
We then compare them with recent 
experiments and numerical calculations \cite{disorder1,disorder1a,disorder3, 
modugno}. In the experiments \cite{disorder2, sch05}, the BEC was prepared within the trapping and random potentials, but its expansion has been studied while switching off these potentials. This led to the observation of sharp fringes in the resulting density due to interference between
different parts of the condensate. These conditions therefore differ from the case we consider. 

We recall that our disorder is
characterized by its strength $s$ in units of the
chemical potential $\mu$ and by the length scale $\lambda_c$ of its
spatial variations. The latter quantity is analogous to the correlation length of disorder defined in \cite{disorder3}. It is
important to stress that in the Thomas-Fermi regime, the time evolution is very sensitive to the existence of potential barriers of height larger than the chemical potential $\mu$. If such a barrier exists, say at a point $z_0$, then we observe that the density $\rho_\omega (z)$ vanishes 
for $z \geq z_0$ at any subsequent times so that the cloud
becomes spatially localized. Then, the average parameter $\langle
s \rangle$ is not anymore relevant since it may be smaller than
unity although some barriers may be larger than $\mu$. We thus need to
characterize the disorder by means of higher moments. For a smooth
enough probability distribution of the random variable $V_m
\omega^c$, which is the case we consider, it is enough to consider
the variance $\delta \omega^c$ defined by $\delta \omega^c =
\left( \langle \left( \omega^c \right)^2 \rangle - \langle \omega^c
\rangle^2 \right)^{1/2}$ and the parameter \beq \delta s = {V_m \over \mu}
\delta \omega^c \label{variance} \eeq which sets the width of
the distribution of potential barriers. In some of the cases we
consider, the peak height of the disordered potential is twice as high as $V_m \delta_{\omega_c}$.  A specific feature of one-dimensional disorder is that it is always very strong
in contrast to higher dimensional systems for which the cloud may
always find a way to avoid large
potential barriers thus making effects of disorder comparatively
weaker. 

\begin{figure}[ht]
\centerline{ \epsfxsize 10cm \epsfysize 7cm \epsffile{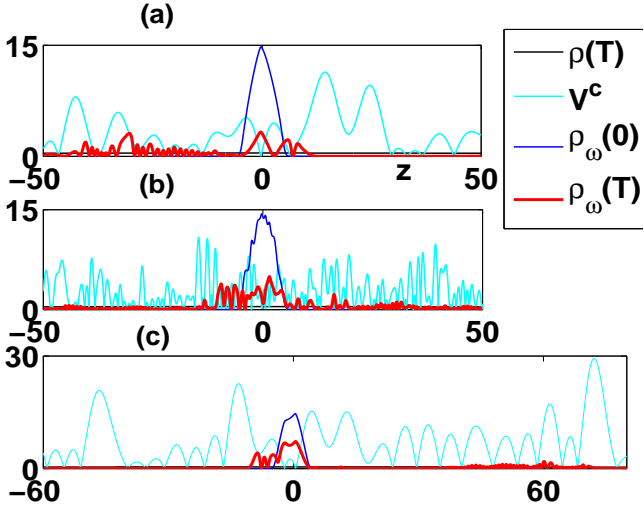} }
%\epsfbox{TFfig_2a.eps}
%\vskip1.0pc
\caption{\em (a) Plot of the density profile $\rho_{\omega}=|\phi_\omega (z)|^2$
for two different times, $T=0$ and $T=25/\omega_z$ and for a slowly
varying disorder such that $\lambda_c \approx 12 \xi$. Both the average
strength of the potential $\langle s \rangle = 0.104$, and
$\delta s = 0.078$ are  kept much below unity and $\mu=30$.The variables plotted in the vertical 
axis are indicated on the right hand side of the upper figure. The density
profile in the absence of disorder at $T=25/\omega_z$ is shown in all three
figures. The cloud extends almost uniformly. The horizontal axis label is 
indicated in the first figure only.
(b) Time evolution of the density for a stronger spatial variation
of disorder  such that $\lambda_c \approx 2 \xi$. $\langle s
\rangle = 0.095$ and  $\delta s  = 0.076$ are  kept well below
unity ($\mu=30$). 
 (c)Time
evolution of the density for a stronger disorder 
characterized by $\langle s \rangle =0.23$, $\delta s
= 0.185$  and $\lambda_c=12 \xi$. $\alpha_{1d}=1$ for all three plots}
\label{fig:TF-den-time1}
%%\end{center}
\end{figure}

In Fig.\ref{fig:TF-den-time1} we have plotted
the density profile $\rho_{\omega} (T)$ after a time $T$ for different spatial variations of 
disorder. For instance, Figs.3{\it (a)} and {\it (b)} compare cases with 
different values of $\lambda_c$ but keeping the disorder strength 
$\delta s \ll 1$, almost unchanged. In 
contrast,  Fig.\ref{fig:TF-den-time1}{\it (a)} and {\it (c)}  
display time evolutions for two disorder
 potentials having the same value of $\lambda_c$, but different strengths 
$\delta s$. In Fig.\ref{fig:TF-den-time1}({\it c}), one of 
the potential well has a height almost equal to $\mu$. 
A first general observation is that for smaller values of $\lambda_c$, namely 
for stronger spatial fluctuations, the spatial expansion of the cloud is more inhibited, so that the main part of the cloud remains localized in 
finite regions that depend on the local landscape of the disordered potential. 
This trend is clearly apparent in Fig.\ref{fig:TF-den-time1}. On the other hand, for small 
values of $\delta s$, small amplitude density fluctuations extend far apart from 
the initial point whereas for large values of $\delta s$, density fluctuations 
do not extend beyond the closest barrier of height larger than $\mu$.

%--------------------------------------------------------------------------------
%\begin{figure}[h]
%\begin{center}
%\vspace*{13pt} \leavevmode \epsfxsize0.75\columnwidth
\begin{figure}[ht]
\centerline{ \epsfxsize 10cm \epsfysize 6cm \epsffile{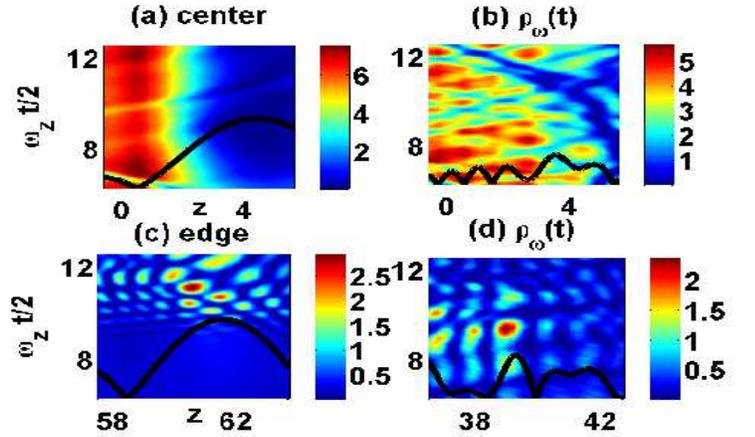} } \caption
{\em
(a) and (c) Time evolution of the density at the center and at 
the edge for a strong disorder ( $\langle s \rangle = 0.23$, 
$\delta s = 0.185$ and $\lambda_c = 12 \xi$) identical to the one used in 
 Fig.\ref{fig:TF-den-time1}{\it (c)}.  
(b) and (d): Time
evolution of the density at the center and at the edge for a disorder characterized by $\langle s \rangle =0.095$,  $\delta s = 0.076$  and $\lambda_c=2 \xi$, identical to the one used in
 Fig.\ref{fig:TF-den-time1}{\it (b)}.
$\mu=30$ and  $\alpha_{1d}=1$ for all figures. The horizontal, vertical 
and color-axis are the same for all three plots and are shown in alternative 
pair of figures. The black line in each figure 
shows the corresponding disordered 
potential. The potential is rescaled and its origin is shifted
by the same amount in all figures to fit it in the size.}
\label{fig:TF-den-time2}
%\end{center}
\end{figure}

%--------------------------------------------------------------------------------
It has been pointed out in \cite{disorder3} that, 
during the time evolution of the Thomas-Fermi cloud, 
its center and its edge 
behave in a different way.  After the trap potential is released, the density 
peak at the center, 
that corresponds to the highest value of the stationary density, gets lowered 
at an initial stage
of the expansion. The interaction energy being larger than the kinetic energy,
the density profile near the center still closely follows a Thomas-Fermi shape,
 but with a reduced chemical potential. 
The spatial variation of  density fluctuations corresponds approximately to 
that of the disordered potential (that is of the order of
$\lambda_c$).  At the edges of the cloud, the density is lower so that the 
kinetic energy term takes over the interaction term and it is almost equal 
to the chemical potential $\mu$ of the
condensate at $t=0$. Thus, the characteristic scale of spatial variations of 
density fluctuations at the edges of the cloud is the coherence length $\xi$
which is smaller than $\lambda_c$. This is displayed in 
Fig.\ref{fig:TF-den-time2} which depicts the time
evolution at the center ({\it (a)} and {\it (b)}) and at the edge ({\it (c)} 
and {\it (d)}). 
In Fig.\ref{fig:TF-den-time2} {\it (a)}, the center of the time evolved cloud
follows the potential landscape and varies on a much larger length 
scale than the edge of the cloud. The other limit shown in 
Fig.\ref{fig:TF-den-time2} ({\it (b)} and  {\it (d)}), displays relatively less difference between spatial 
variations of density fluctuations at the center and at the edges of the 
cloud, since $\lambda_c \simeq \xi$.

In addition, Figs.\ref{fig:TF-den-time2} describe how the matter wave
behaves close to a single potential barrier, at the center and at the edge
of the cloud. The shape of a typical potential is controlled by 
changing $\lambda_c$. In Figs.\ref{fig:TF-den-time2} {\it (a)} and {\it (b)}, 
it is shown how the central cloud becomes localized due to the presence of a potential barrier. The density 
modulation is driven by the  the local potential landscape, rather than 
any interference effect. 
It has been pointed out
in \cite{disorder1a,modugno}
that the height of a single defect should vary like the energy $E$
of the incoming wavepacket over a distance short compared to its 
de Broglie wavelength in order to allow for quantum effects to dominate 
and eventually lead to Anderson localization. The potential 
used in our computation, generally does not satisfy this criterion. 
To satisfy it, one needs a disorder  
with higher $\delta s$ and lower $\lambda_c$. However under 
such conditions, the mean-field 
Gross-Pitaevskii approximation is questionable and the use of a discrete non-linear 
Schr\"odinger equation will be more appropriate.
   
%----------------------------------------------------------
\begin{figure}[ht]
\centerline{ \epsfxsize 9cm \epsfysize 8cm \epsffile{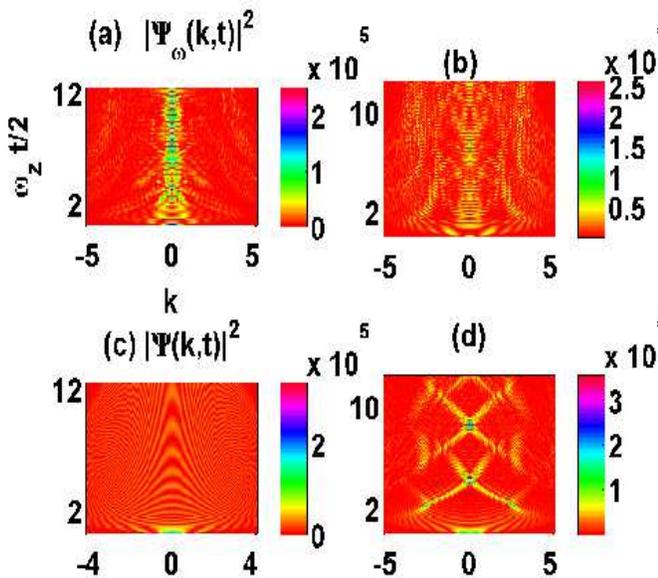} }
 \caption{\em Density evolution in the $k-space$. Horizontal and vertical axis labels are identical for all the plots and are shown in (a).
The color-axis label is indicated in (a) and (c).
The four cases correspond to 
(a) Strong disorder as defined in
 Fig.\ref{fig:TF-den-time1}{\it (c)}. 
 (b) Weaker disorder fluctuating on a smaller length scale as defined in Fig.\ref{fig:TF-den-time1}
{\it (b)}.
(c) No disorder
(d) Optical lattice: $(\langle s \rangle + \delta s)\sin \frac{2 \pi z}
{\lambda_c}$. The values of $\langle s \rangle$, $\delta s$ and $\lambda_c$ are those used
in Fig.\ref{fig:TF-den-time1}{\it (c)}.}
\label{fig:k-space}
%\end{center}
\end{figure}

%----------------------------------------------------------
We have studied in Fig.\ref{fig:k-space} the time evolution of the cloud density in momentum space and compare it to the cases without disorder and in the presence of an optical lattice. To make the comparison easier, we have used
the amplitude of the optical lattice potential corresponding to that, $<s>+\delta s$, considered in Fig.\ref{fig:k-space} {\it (a)}. 
Fig.\ref{fig:k-space} {\it (a)} shows a strong localization in $k$-space for high $\delta s$. This may be  compared to the situation 
of Fig.\ref{fig:k-space}{\it (d)} (optical lattice). This strong localization
occurs around the $k=0$ state.  
On the other hand when disorder fluctuates on a shorter scale $\lambda_c$ (with a smaller $\delta s$), a significant fraction of the density 
still occupies higher momentum states and the corresponding 
localization in momentum space is less pronounced. Thus, an experimental measurement of the momentum spectrum \cite{aspect2} of a  
quasi one-dimensional BEC in a disordered 
waveguide can shed light on the nature of localization
of the cloud.
%==========================================================

%----------------------------------------------------------  
\medskip

\begin{figure}[ht]
\centerline{ \epsfxsize 10cm \epsfysize 7cm \epsffile{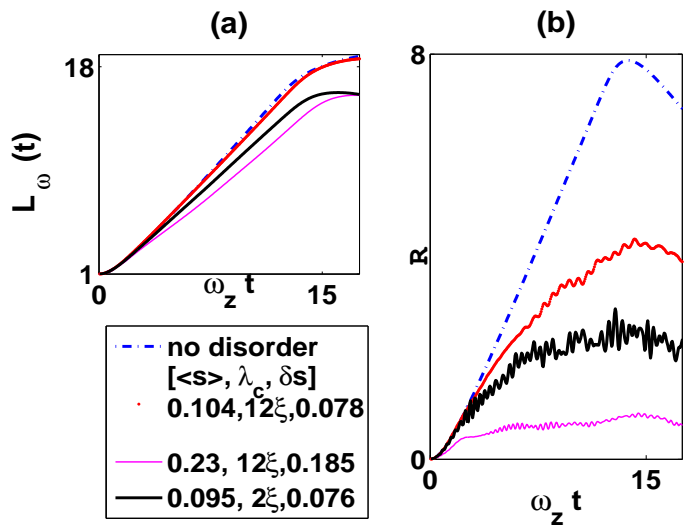} } \caption{\em
(a) Time evolution of the spatial extension $ L_\omega (t)$ of the cloud as defined in
(\ref{spatialext}) for a given configuration of the disordered  potential but for different
strengths $V_m$ and length scales $\lambda_c$. 
 $\langle s \rangle$, 
$\lambda_c$ and $\delta s$ are indicated against  each plot. $L_\omega (t) $ is expressed in units of  its value at
$t=0$. (b) Corresponding
time evolution of the ratio $\cal R$ defined in (\ref{ratioR}). }
\label{fig:TFeratio-size}
%\end{center}
\end{figure}

After studying the time evolution of the density, we now discuss
the time evolution of other properties of the cloud that
characterize the suppression of its expansion.  In Fig.\ref{fig:TFeratio-size}({\it a}), the spatial extension 
$ L_\omega $ for a given configuration, is plotted as a function of the
dimensionless time $\omega_z t$. We 
observe that $ L_\omega (t) $ saturates with time to a value 
which depends on the average strength $\langle s \rangle$ of the disorder. 
 
In order to characterize this saturation, we define the
ratio, denoted by $\cal R$, between the average kinetic and
interaction energies of the cloud, defined by \beq {\cal R} =
2\xi^2 {\int dz  \left( {\partial \phi_\omega \over \partial
z} \right)^2 \over \int dz |\phi_\omega |^4} \, \ . \label{ratioR}
\eeq In the stationary Thomas-Fermi approximation,  the kinetic
energy is almost negligible as compared to the interaction term,
{\it i.e.}, ${\cal R} \approx 0$. As the cloud expands, the interaction
energy gets gradually converted into kinetic energy and this ratio
increases. Finally, it saturates when almost all the interaction
energy is converted into kinetic energy. This shows up in Fig.\ref{fig:TFeratio-size}({\it b}). For a larger disorder,  this increase of the ratio
saturates more rapidly  and the slope of $\cal R$, which indicates 
how fast the interaction energy is converted into kinetic energy,
decreases. Particularly the lowest plot corresponding to a large disorder, shows 
a rapid saturation of $\cal{R}$ due to strong localization in momentum space. 
Since the edge of the cloud involves mostly kinetic
energy, the behavior of $\cal R$ is dominated by the
expansion of the central region. When the expansion is stopped by a potential barrier, the corresponding loss
in kinetic energy is proportional to the height of the potential
barrier. This explains the oscillations of $\cal R$ that appear 
in the presence of disorder.

%----------------------------------------------------------

%----------------------------------------------------------

\section{Gaussian limit}

\begin{figure}[ht]
\centerline{ \epsfxsize 8cm \epsffile{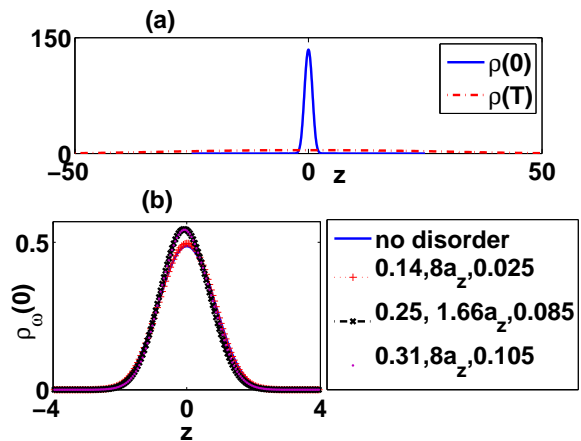}
}\caption{\em (a) Time evolution of the condensate density in the
absence of disorder. $\rho(0)$ is the stationary density and
$\rho(T)$ is the density after a time $T=\frac{25}{\omega_z}$. We
have taken $\mu=2$ and $2\alpha_{1d}=0.01$ (b) Stationary profile
of the condensate density in the presence of disorder. The corresponding 
values of the average disorder strength $\langle s \rangle$ , 
$\lambda_c$ and $\delta s$ are given in the inset.} 
\label{fig:Gauss-den-time1}
%\end{center}
\end{figure}

In this section we study effects of disorder
on bosons that are condensed in the ground state of a harmonic
oscillator potential. In that case, solutions of the
Gross-Pitaevskii equation without disordered potential, are different from those
observed in the Thomas-Fermi limit, and are given by Gaussian profiles
centered at the origin. When the trap potential is released, corresponding time-dependent solutions
 remain  Gaussian but with a larger width, a standard result from
quantum mechanics (see Fig.\ref{fig:Gauss-den-time1}({\it a})).

%----------------------------------------------------------
%\begin{figure}[h]
%\begin{center}
%\vspace*{13pt} \leavevmode \epsfxsize0.75\columnwidth
%\epsfbox{Gaussfig_1.eps} \vskip1.0pc
%----------------------------------------------------------

\begin{figure}[ht]
\centerline{ \epsfxsize 7cm \epsfysize 5cm \epsffile{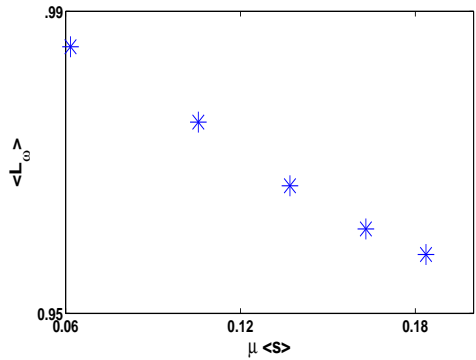} }
\caption{\em Plot of the disorder averaged width $\langle L_\omega \rangle$ of the stationary density profile in the
Gaussian limit as a function of the average disorder strength $\mu \langle s \rangle$.
The width is normalized by its value in the absence of disorder. For a
given value of $\lambda_c$, the width is averaged over $200$
realizations of the potential. The values of $\lambda_c$ are those used  in
Fig.\ref{fig:TFwidth} and, in units of the harmonic oscillator
length $a_z$, they range between $0.55a_z$ (lowest point) and
$5 a_z$ (highest point). We have used $\mu=2$ and
$2\alpha_{1d}=0.01$. } \label{fig:Gausswidth}
%\end{center}
\end{figure}
\subsection{Stationary solutions}

 Like for the Thomas-Fermi regime, stationary solutions of
 the Gross-Pitaevski equation (\ref{gnlse2}) in the presence of both trapping and disorder are characterized by the average strength $\langle s \rangle$ and the length $\lambda_c$. By changing the disorder strength we obtain behaviors  such
as those displayed in Fig. \ref{fig:Gauss-den-time1}({\it b}). 

%----------------------------------------------------------
\begin{figure}[ht]
\centerline{ \epsfxsize 10cm \epsfysize 10cm \epsffile{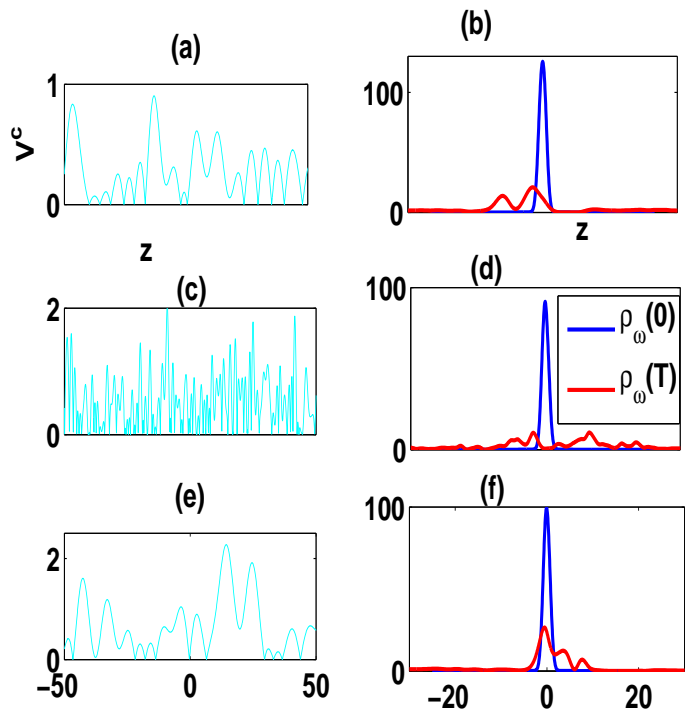} }
%\vskip1.0pc
\caption{\em {\it Left:}(a,c,e) Plot of the disordered potential.
The potentials in (a,e) vary on the same scale, whereas the
potential in (c) varies on a  smaller scale. The corresponding values of $\langle s \rangle$, 
$\lambda_c$ and $\delta s$ are respectively $(0.14, 8a_z, 0.025)$, $(0.25, 1.66a_z, 0.0846)$ and
$(0.31, 8a_z, 0.105)$. 
%The length scale of variation
%is expressed in terms of the harmonic oscillator length along the $z$ axis, $a_z$.
{\it Right:}(b,d,f) Time evolution of the density corresponding
to the potentials plotted on the left. The initial stationary
value of the density is denoted by $\rho_{\omega}(0)$ and $\rho_{\omega}(T)$ is computed at the time $T = 25/\omega_z$. We have used $\mu=2$ and
$2\alpha_{1d}=0.01$.} \label{fig:Gauss-den-time2}
%\end{center}
\end{figure}
%----------------------------------------------------------

Since interaction effects are negligible in the Gaussian limit,
the characteristic length of density variations  is set by  the
harmonic oscillator length $a_z$, and not by the coherence length
$\xi$ as before, the latter being very large in that case. In this regime dominated by confinement, 
we observe that  the shape of the density profile depends weakly on disorder in contrast to the
Thomas-Fermi limit, for which this profile follows the variations
of the disorder. This is particularly apparent in Figures 
\ref{fig:Gauss-den-time2}({\it c,d}),  where disorder varies over
a length scale smaller than the width of the density profile without
leading to fluctuations of this profile.

The density profile is well approximated by a off-centered Gaussian shape, 
\beq
\rho_\omega (z)=A \exp \left( -\frac{(z-z_0)^2}{ L_\omega ^2} \right) \, ,
\eeq
for a large range of disordered potentials (see Figure \ref{fig:Gauss-den-time1}({\it b})). The amplitude $A$ and the width $ L_\omega $ are
related to each other through the normalization. The average width $\langle L_\omega \rangle$ is a decreasing function of  the disorder strength $\mu \langle s \rangle$ defined by (\ref{parameters}) as represented in
Fig.\ref{fig:Gausswidth}. Thus, the net effect of disorder  is to spatially
localize the bosons inside a narrower Gaussian.

%----------------------------------------------------------
%\begin{figure}[h]
%\begin{center}
%\vspace*{13pt} \leavevmode \epsfxsize0.75\columnwidth
%\epsfbox{Gaussfig_1a.eps}
%\vskip1.0pc

%----------------------------------------------------------

\subsection{Time dependent solutions}

The behavior of the stationary condensate density profile in the presence of
disorder in the Gaussian limit differs from the one obtained in the Thomas-Fermi limit. This difference  shows up also in the time evolution of the density  of the
cloud after switching off  the trapping potential. The short time expansion of the Thomas-Fermi cloud strongly depends on disorder, whereas in the Gaussian case, it does not.
Moreover, in contrast to the Thomas-Fermi case, the zero point motion of the bosons is appreciable. The time evolution of the condensate density after switching off the
trap is presented in Fig. \ref{fig:Gauss-den-time2} for different strengths of disorder.

%Notice that the Gaussian limit can be achieved
%either by increasing the harmonic trap frequency or by decreasing the $s$-wave scattering length.
%In the first case, the real time expansion is much smaller
%for the Gaussian cloud and may not be enough for observing any disorder induced trapping. Here we assume
%that Gaussian regime can be achieved following the second way, namely by changing $s$-wave scattering length.

%----------------------------------------------------------
\begin{figure}[ht]
\centerline{ \epsfxsize 10cm \epsfysize 6cm \epsffile{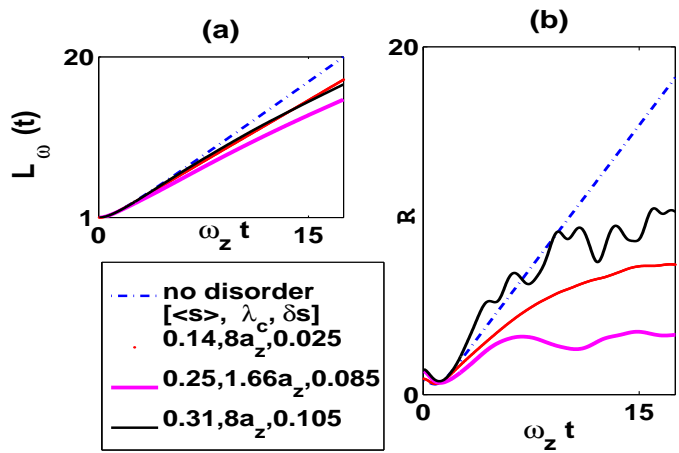}
}\caption{\em (a) Time evolution of the spatial extension $ L_\omega (t) $ of the cloud for disorders of  different
strengths $V_m$ and length scales of variation $\lambda_c$. The disordered 
potentials are those used in Fig.
\ref{fig:Gauss-den-time2} and the corresponding parameters are 
 indicated ($\langle s \rangle$, $\lambda_c$, $\delta s$)
in the inset. $ L_\omega (t) $ is
expressed in units of its value at $t=0$. 
 (b) Corresponding time evolution of the ratio $\cal R$ defined in (\ref{ratioR}). } \label{fig:Gausseratio-size}
%\end{center}
\end{figure}
%----------------------------------------------------------

We first notice that on the same time scale, the density at the center of the cloud decreases more rapidly than for the Thomas-Fermi case (Figs.\ref{fig:TF-den-time1} and \ref{fig:TF-den-time2}). This results from  the non negligible kinetic energy of  a Gaussian cloud and the weaker interaction between bosons. Figure \ref{fig:Gausseratio-size} displays the time evolution of
the average spatial extension $\langle L_\omega \rangle$ of the cloud defined by (\ref{spatialext})
and the ratio of the average kinetic and interaction energies defined in (\ref{ratioR}). These two figures outline the
difference between Thomas-Fermi and Gaussian time evolutions in
the presence of  disorder. The spatial extension in
Fig.\ref{fig:Gausseratio-size}({\it a}) does not show any
saturation over comparable time scales, though it grows at a
lesser rate with increasing the strength of disorder. Correlatively, the
ratio $\cal R$ in Figure \ref{fig:Gausseratio-size}({\it b})
grows at a much faster rate and it takes a longer time to
saturate. We can summarize these observations by saying that
though the cloud expansion is indeed prevented by the disorder
potential in the Gaussian regime, the suppression is weaker  than in the Thomas-Fermi regime and it happens on longer time
scales.
\section{Soliton solutions for an attractive 
one-dimensional  Bose-Einstein condensate}

Having discussed the behavior of repulsive  interacting bosons in the presence of disorder,  we
now turn to the case of an attractive solitonic condensate in similar situations.
As we shall see, the change of the nature of the interaction modifies the behavior
of the soliton solution with disorder as compared to the previous cases of Thomas-Fermi and Gaussian condensates. In contrast to Eq.(\ref{dimnlse}) describing a repulsive interaction, Eq.(\ref{ANSEE1}) involves one free parameter only $(\alpha_{1d} = -1)$. As we have already mentioned, a change in $\alpha_{1d}$ only redefines the width of the soliton proportional to $1/ \sqrt{\mu}$. In what follows, the width is always kept less than $\xi$.

\subsection{Stationary profiles}

We start with the study of the stationary solutions of Eq.(\ref{ANSEE1}) with the addition of a random potential, namely,
\beq
- \mu \phi (z) + \partial_z^2 \phi(z) +V_d \phi(z) + 2 |\phi|^2 \phi =0 \, \ .
\eeq
 It is important to notice that, in contrast
with previous cases, there is no  trapping potential, so that in
the absence of disorder, the solution is invariant under
translations. Numerically, we start with a randomly chosen initial
guess which, once iterated,  gives a  solution located around the
initial trial function. The overall shape of the stationary
solution turns out to be independent of disorder, meaning that
this shape can still be fitted with a function
%----------------------------------------------------------
\begin{figure}[ht]
\centerline{ \epsfxsize 7cm \epsffile{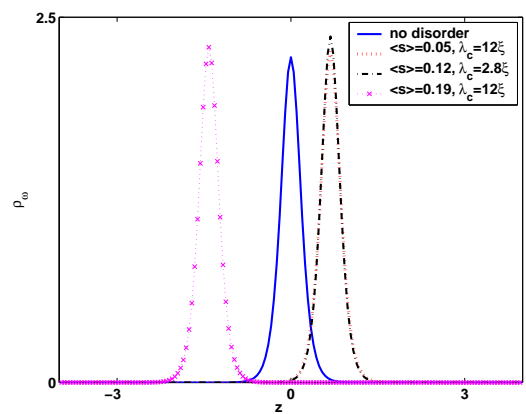} }\caption{\em
Stationary density profile of a bright soliton in the presence
of disorder. The chemical potential is $\mu=-20$ and
$\alpha_{1d}=-1$. The disordered potential is characterized by $|s|$
(since $\mu <0$) and $\lambda_c$. $<s>$ is the average of $|s|$. }
\label{fig:soliton-profile}
%\end{center}
\end{figure}
%----------------------------------------------------------
of the type $A_s sech(B_s(z - z_0))$, where $A_s$ and $B_s = 1/  L_\omega $ are
respectively  the amplitude and the inverse width of the soliton.
This feature appears clearly  in Fig.(\ref{fig:soliton-profile})
where the profile of the bright soliton has been plotted for
several realizations of the potential. But, both the width and the amplitude depend on disorder
as shown in Fig.(\ref{fig:SLwidth}) which displays the behavior
of the width for an increasing strength of  disorder.  We have
also checked the dependence upon length scales  $\lambda_c$. Those features look similar to those obtained in the Gaussian limit. But they are essentially different.
Whereas the soliton profile results from the comparison between
kinetic and negative interaction energies, the Gaussian profile is
obtained from the comparison between kinetic and confinement
energies. This difference will manifest itself in the time
evolution of the solitonic condensate.
%----------------------------------------------------------
\begin{figure}[ht]
\centerline{ \epsfxsize 9cm \epsfysize 7cm 
\epsffile{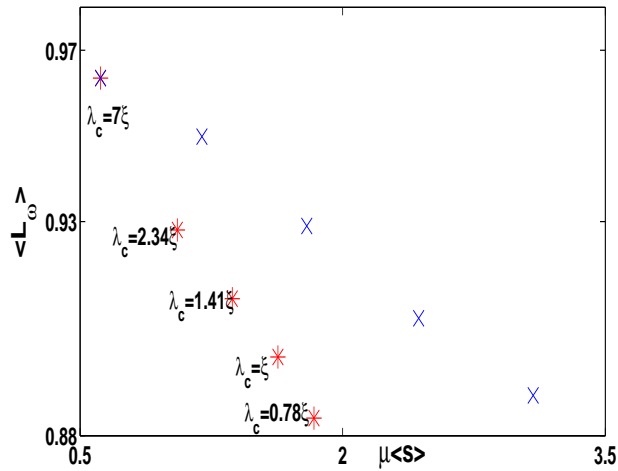} } \caption{\em Disorder averaged width $\langle L_\omega \rangle$ of the stationary profile of a soliton as a
function of the strength of disorder. For one set of points
($\times$), the strength $V_m$ of the disordered potential is
increased keeping fixed the spatial scale of variation $\lambda_c$. For the
other set of points ($*$), $\lambda_c$ is lowered which corresponds to stronger fluctuations of the disorder, while $V_m$ is kept fixed. $\langle L_\omega \rangle$ is expressed in
units of its value in the absence of disorder and, for each case,
it is averaged over $200$ realizations of the potential.
The average potential is characterized by $|s|$ since
$\mu < 0$. The values of $\lambda_c$ are indicated in the figure. We have taken $\mu=-20$ and $2\alpha_{1d}=-1$. }
\label{fig:SLwidth}
%\end{center}
\end{figure}
%----------------------------------------------------------

\subsection{Time-dependent solutions}

\begin{figure}[ht]
\centerline{ \epsfxsize 9cm \epsfysize 10cm \epsffile{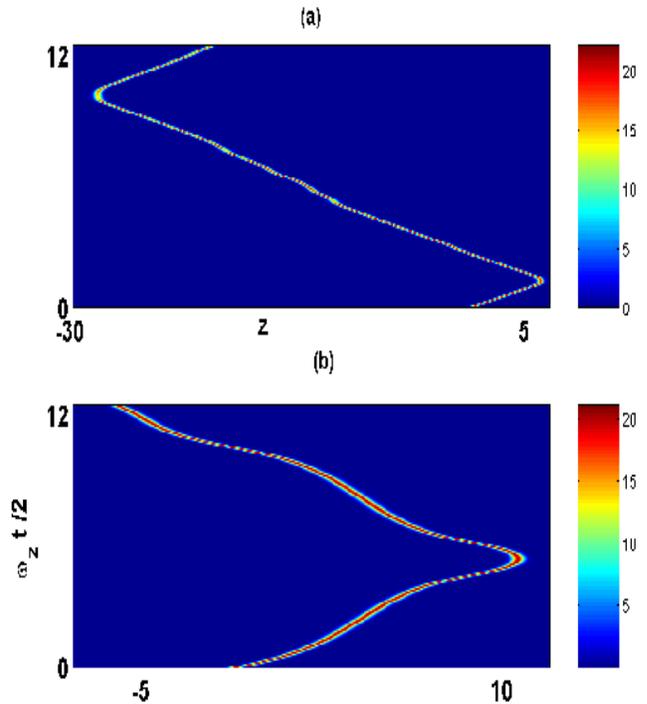} }\caption{\em
Time evolution of a boosted soliton in the presence of disorder.
The chemical potential is $\mu=-20$, the dimensionless velocity at $t=0$ is
$V_s = 5$ and $\alpha_{1d}=-1$. The disorder potential is characterized
by $|s|$ (since $\mu <0$) and $\lambda_c$. (a) Fastly varying
disorder with $<s>=0.12$, $\delta s=0.162$ and $\lambda_c=2.8 \xi$.
(b) Stronger but slowly varying disorder with $<s>=0.19$,
$\delta s =0.372$ and $\lambda_c=12 \xi$.}
\label{fig:soliton-motion2}
%\end{center}
\end{figure}

We now study the time evolution of the stationary solutions
obtained previously, and not initial solutions given by (\ref{ANLSE}) unlike the case
considered in \cite{KIV90}.  To this purpose, we first boost the
soliton by giving it a finite (dimensionless) velocity  $V_s=5$.  In the absence of disorder, the soliton travels a distance $z=V_s T$ over a time $T$ without any change 
in its density profile. In the presence
of a weak and smooth enough disorder, we
observe that the soliton propagates retaining its initial ($t=0$) shape, over 
distances comparable to the non disordered case.  A weak disorder
potential has thus a negligible effect on the soliton motion. For a stronger disorder strength  ({\it
i.e.}, for a smaller value of $\lambda_c$ and a larger value of
$<s>$), the time behavior is displayed in Figures \ref{fig:soliton-motion2}({\it a})
and ({\it b}). In both cases, the soliton behaves classically and
it becomes spatially localized, {\it i.e.} that it bounces back from
high potential barriers typically higher than the kinetic energy.
However, we do not observe a significant change in the shape of the
soliton. Its width fluctuates as the soliton travels through the
disordered potential and bounces back and forth. When 
the strength of disorder is higher, 
the soliton motion is clearly not linear (Figure \ref{fig:soliton-motion2}({\it b})). 
This kind of motion can be qualitatively explained by considering the 
soliton as a massive classical particle of mass $m N$, where $m$ and $N$ are 
respectively the mass and the number of atoms in the condensate. Deviations 
from the linear motion result from the spatially varying force exerted on the 
soliton by the disorder potential. This kind of description is valid as long 
as the disorder potential remains smooth over the width of the soliton. 
Similar behaviors have been discussed in the context of soliton chaos 
in spatially periodic potentials \cite{martin}, although the physical origin 
is different from the case discussed here. With the present stage of experiment
\cite{soliton02}, such a behavior can be verified by studying the
time evolution of a bright soliton in an optical speckle pattern.

%----------------------------------------------------------

%----------------------------------------------------------
%----------------------------------------------------------

%----------------------------------------------------------
%\begin{figure}[ht]
%\centerline{ \epsfxsize 8cm \epsffilesolfig_4.eps} }
% \caption{\em  (a)Time evolution of the
%soliton width. The disorder potential and the other
%parameters are  those used in the previous plots. The width
%is given in units that  of the static soliton  (b) Corresponding
%time evolution of the soliton peak. }
%\label{fig:soliton-width-peak}
%%\end{center}
%\end{figure}
%----------------------------------------------------------

\section{Conclusion}

We have performed a detailed numerical investigation of stationary solutions and time evolution
of one dimensional Bose-Einstein condensates in the presence of a random  
potential. Stationary solutions which correspond either to
the attractive interaction bright soliton or to repulsive interaction Gaussian matter waves with repulsive interactions in the regime where confinement dominates, behave in a 
qualitatively 
similar way. In contrast,
the stationary solutions that correspond to a repulsive interacting
Thomas-Fermi condensate, depend strongly on the strength of disorder and 
on its spatial scale of variations.

The time evolution of stationary solutions depends also significantly
on the regime we consider. Although transport gets inhibited both
for the attractive and repulsive interaction, this occurs in a
very different way. For the repulsive case the center and the edge
of the cloud behave differently and both are ultimately localized
in a deep enough potential well. In the interaction dominated
Thomas-Fermi regime,  the main part of the cloud remains localized
and edges that correspond to low densities and correlatively
weaker interactions, propagate further away.  A study of the corresponding momentum
distribution of the cloud indicates a stronger localization of the matter
wave in low momentum states for an increasing strength of the disorder potential. 
On the other hand, a
moving bright soliton behaves very much like a single particle and it bounces back from a steep potential with its motion reversed. This behavior of a bright soliton may be contrasted against
the behavior of a dark soliton in the presence of disorder which has been
investigated recently \cite{NBNP05}.

\medskip

For the values of the disorder strength and the non-linearity we have considered,
we observe a behavior of solutions of the Gross-Pitaevskii
equation that are mostly driven by the non-linearity, {\it i.e.}, by
interactions. Disorder plays mostly the role of a landscape within
which a classical solution evolves in time. We did not observe,
for the relatively large range of disorder and interaction
parameters we have considered, a behavior close to Anderson
localization, namely where spatially localized solutions result
from interference effects. 
Since disorder is expected to
be stronger in one-dimensional systems, we may conclude that, for
the currently accessible experimental situations, Anderson
localization effects will not be observable \cite{disorder1a,
sch05, SR94} due to the strength of the interaction term.
Alternative 
setups are thus required in order to observe quantum localization
of matter waves, having weak or zero interaction ({\it e.g.}, by
monitoring Feschbach resonances \cite{GC04}).

The signature of Anderson localization in the nonlinear transport of a BEC in a wave-guide geometry has been studied in \cite{paul}. There, the transmission coefficient has been shown to be exponentially decreasing with the system size below a critical interaction strength. But the different types of disorder and of the matter wave density at $t=0$, make a direct comparison 
with these results difficult.
%We have nevertheless observe interesting and unexpected effects of a disorder potential.
%In the Thomas-Fermi  limit,

\section{Acknowledgments}
%E. Akkermans  and S. Gosh  thank B. Shapiro and S. Fishman for helpful
%discussion and their interest in this work.
S. Ghosh thanks Hrvoje Buljan for his help in numerical
computation. This research is supported in part by the Israel
Academy of Sciences and by the Fund for Promotion of Research at
the Technion.


\begin{thebibliography}{99}
\bibitem{Ander58}P. W. Anderson, Phys. Rev. {\bf 109}, 1492
(1958)
\bibitem{parodi} M. Belzons, E. Guazzelli and O. Parodi, J. of Fluid Mech. {\bf 186}, 539
(1988); M. Belzons, P. Devillard, F. Dunlop, E. Guazzelli, O.
Parodi and B. Souillard, Europhys. Lett. {\bf 4}, 909 (1987)
\bibitem{kergomard} C. D\'epollier, J. Kergomard and F. Lalo\"e, Ann. Phys. (France) {\bf 11}, 457 (1986)
\bibitem{amaynard} E. Akkermans and R. Maynard, J. Physique (France) $\bf 45$, 1549 (1984) 
\bibitem{Gang479}E. Abrahams, P.W. Anderson, D. C. Licciardello,
and T. V. Ramakrishnan, Phys. Rev. Lett. {\bf 42}, 673 (1979).
\bibitem{genack} A.A. Chabanov, M. Stoytchev and A.Z. Genack, Nature, {\bf 404}, 850 (2000)
\bibitem{lag} D.S. Wiersma, P. Bartolini, A. Lagendijk and R. Righini, Nature, {\bf 390}, 671 (1997)
\bibitem{maret} F. Scheffold, R. Lenke, R. Tweer and G. 
Maret, Nature  {\bf 398}, 207 (1999)
\bibitem{review1} B. Kramer and A. MacKinnon, Rep. Prog. Phys. {\bf 56}, 1469 (1993).
\bibitem{review2}P.A. Lee and T.V. Ramakrishnan, Rev. Mod. Phys. {\bf 57}, 287 (1985)
\bibitem{am} E. Akkermans and G. Montambaux,
\textit{Physique m\'esoscopique des \'electrons et des photons},
(Paris, EDP Sciences 2004). English translation (Cambridge
University Press)
\bibitem{josaam} E. Akkermans and G. Montambaux, J. Opt. Soc. Am.  B {\bf 21}, 101 (2004)
\bibitem{DS86}P. Devillard and B. Souillard, J. of Stat. Phys. {\bf 43}, 423
(1986).
\bibitem{DR87}B. Doucot and R. Rammal, Eur. Phys. Lett. {\bf 3}, 969 (1987);
 {\it ibid} J. Phys.   (Paris), {\bf 48}, 527 (1987)
\bibitem{KIV90}Y.S. Kivshar, S.A. Gredeskul, A. Sanchez and L. Vazquez,
Phys. Rev. Lett. {\bf 64}, 1693 (1990).

\bibitem{disorder1}J.E. Lye, L. Fallani, M. Modugno, D. Wiersma, C. Fort and M. Inguscio, Phys. Rev.
Lett. {\bf 95}, 070401 (2005).
\bibitem{disorder1a} C. Fort, L. Fallani, V. Guarrera,
J. Lye, M. Modugno, D. S. Wiersma and M. Inguscio, Phys. Rev.
Lett. {\bf 95}, 170410 (2005).
\bibitem{disorder2}P. Kruger, L.M. Andersson, S. Wildermuth, S. Hofferberth, E. Haller, S. Aigner, S. Groth,
I. Bar-Joseph, J. Schmiedmayer, arXiv:cond-mat/0504686.
\bibitem{disorder3}D. Cl\'ement, A. F. Var\'on ,
 M. Hugbart, J. Retter, P. Bouyer, L. Sanchez-Palencia, D.M. Gangardt, G. V. Shlyapnikov, A.
 Aspect, Phys. Rev. Lett. {\bf 95}, 170409, (2005).
 \bibitem{sch05}T. Schulte, S. Drenkelforth, J. Kruse, W. Ertmer,
 J. Arlt, K. Sacha, J. Zakrzewski., and M. Lewenstein, Phys. Rev.
 Lett. {\bf 95}, 170411 (2005).
\bibitem{Dam03}B. Damski, J. Zakrzewski, L. Santos, P. Zoller, and M. Lewenstein
Phys. Rev. Lett. {\bf 91}, 080403 (2003) and R. Pugatch, N. Bar-Gil, N. Katz, E Rowen and N. Davidson, arXiv:cond-mat/0603571.
\bibitem{SKSZL04}A. Sanpera {\it et al.}, Phys. Rev. Lett. {\bf 93}, 040401 (2004).
\bibitem{Wang04} D.W. Wang, M.D. Lukin, and E. Demler,
 Phys. Rev. Lett. {\bf 92}, 076802 (2004).
\bibitem{GC04}U. Gavish and Y. Castin, Phys. Rev. Lett. {\bf 95},
020401 (2005).
\bibitem{PVC05}B. Paredes, F. Verstraete and J.I. Cirac, Phys. Rev. Lett. {\bf 95}, 140501 (2005).
\bibitem{NBNP05}N. Bilas and N. Pavloff, Phys. Rev. Lett. {\bf 95}, 130403 (2005).
\bibitem{fallani} L. Fallani, J. Lye, V. Guarrera, C. Fort  and M. Inguscio,  arXiv:cond-mat/0603655.
\bibitem{modugno} M. Modugno, Phys. Rev. A {\bf
 73}, 013606 (2006).
\bibitem{paul}  T.Paul, P. Leboeuf, N. Pavloff, K. Richter and P. Schdagheck,
Phys. Rev. A {\bf
 72}, 063621 (2005).
 \bibitem{GWSS05}H. Gimperlein, S. Wessel, J. Schmiedmayer, and L.
 Santos, Phys. Rev. Lett. {\bf 95}, 170401 (2005).
 \bibitem{SR94}K. G. Singh and D. S. Rokshar, Phys. Rev. B {\bf
 49}, 9013 (1994).
\bibitem{Olshanii98}M. Olshanii, Phys. Rev. Lett. {\bf 81}, 938 (1998).
\bibitem{pita} L. Pitaevskii and S. Stringari, {\em Bose-Einstein condensation}, Oxford (2003)
\bibitem{PSW00}D.S. Petrov, G.V. Shlyapnikov and J.T.M. Walraven, Phys.Rev. Lett.
{\bf 85}, 3745 (2000) and D.M. Petrov, D.M. Gangardt and G.V.
Shlyapnikov, J. Phys. IV France 1 (2004).
\bibitem{TG36}L. Tonks, Phys. Rev. {\bf 50}, 955 (1967); M. D.
Girardeau, J. Math. Phys. {\bf 1}, 516 (1960); C. N. Yang and C.
P. Yang, J. Math. Phys. {\bf 10}, 1115 (1969).
\bibitem{soliton02}L. Khaykovich {\it et al.}, Science {\bf 296},
1290 (2002); K. E. Strecker {\it et al.}, Nature {\bf 417}, 150
(2002) and B. Eiermann {\it et al.} Phys. Rev. Lett. {\bf 92},
230401 (2004).
\bibitem{CC02}L.D. Carr and Y. Castin, Phys. Rev. A {\bf 66} 063602 (2002).

\bibitem{book1}C. Sulem and P. L. Sulem, {\it The Nonlinear Schr\"odinger Equation Self-Focusing and Wave Collapse}, Springer, New-York (1999).

\bibitem{P76}V.I.Petviashvili, Sov. Plasma Phys. {\bf 2}, 257 (1976).

\bibitem{AM03}M.J. Ablowitz and Z. H. Musslimani, Physica D, {\bf 184}
276 (2003).
\bibitem{AM04}Z.H. Musslimani and J. Yang, J. Opt. Soc. Am. B {\bf 21}, 973,
(2004).
\bibitem{AM05}M.J. Ablowitz and Z. M. Musslimani Opt. Lett. {\bf 30}, 2140 (2005).
\bibitem{BJM03}W. Bao, D. Jaksch and P.A. Markowich, J. Comput. Phys. {\bf 187}, 318 (2003)
\bibitem{aspect2}S. Richard {\it et al.}, Phys. Rev. Lett. {\bf 91}, 010405 
(2003). 
\bibitem{martin} R. Scharf and A.R. Bishop, Phys. Rev. A {\bf
 46}, R2973 (1992) and  A.D. Martin, C.S. Adams and S.A. Gardinar, arXiv:cond-mat/0604086.

\end{thebibliography}
\end{document}